\documentclass[11pt]{article}

\newcommand{\blind}{0}

\usepackage{amsmath}
\usepackage{graphicx}
\usepackage{enumerate}
\usepackage{enumitem}
\usepackage{natbib} %comment out if you do not have the package
\usepackage{url} % not crucial - just used below for the URL

\graphicspath{{figures/}}
\usepackage{amsfonts, amsthm, latexsym, amssymb}
\usepackage{dsfont}
\usepackage{array}
\usepackage{float}
\usepackage{tikz}
\usepackage{multirow}
\usepackage{xcolor}
\usepackage{lineno}
\usepackage{pgfplots}
\pgfplotsset{compat=1.17}
\usepackage{hyperref}

\usepackage{authblk} 
\usepackage{booktabs}
\usepackage[utf8]{inputenc}
\usepackage[english]{babel}
\usepackage{csquotes} 
\usepackage{graphicx}
\usepackage{subcaption}
\usepackage{algpseudocode}
\usepackage{tikz}
\usetikzlibrary{positioning, arrows.meta}
\usepackage[ruled,vlined]{algorithm2e}

\DeclareMathOperator*{\argmin}{\arg\!\min}

\begin{document}

%%%%%%%%%%%%%%%%%%%%%%%%%%%%%%%%%%%%%%%%%%%%%%%%%%%%%%%%%%%%%%%%%%%%%%%%%%%%%%
\def\spacingset#1{\renewcommand{\baselinestretch}%
{#1}\small\normalsize} \spacingset{1}
%%%%%%%%%%%%%%%%%%%%%%%%%%%%%%%%%%%%%%%%%%%%%%%%%%%%%%%%%%%%%%%%%%%%%%%%%%%%%%

		\if0\blind
  {
\title{Spatio-Temporal Gaussian Process for Building Terrain-Incorporating Wind Power Curves}
%\subtitle{Your subtitle} % Uncomment if you have a subtitle

% Define command for short article title and author last names
%\newcommand{\shortarticletitle}{A Short Title}
\newcommand{\authorlastnames}{Chokhachian, Joseph, Ding}
\author{Ahmadreza Chokhachian, V. Roshan Joseph and Yu Ding\thanks{Corresponding author: Yu Ding, yu.ding@isye.gatech.edu.} \\
H. Milton Stewart School of Industrial and Systems Engineering\\
Georgia Institute of Technology
}

}\fi
		\if1\blind
		{
\title{Spatio-Temporal Gaussian Process for Building Terrain-Incorporating Wind Power Curves}
			\author{Author information is purposely removed for double-blind review}
   \bigskip
			\bigskip
			\bigskip
}\fi

\date{\small \today}
\maketitle
\begin{abstract}
Accurate modeling of wind turbine power curves is crucial for optimal wind farm operation. Nearly all existing power curve models focus on temporal variables such as wind speed and temperature while overlooking the influence of terrain covariates, which governs inflow wind conditions and thus also affects wind power production. This paper proposes a nonparametric spatio-temporal Gaussian process model that integrates temporal environmental covariates with spatial terrain features. The model falls in the category of spatial-temporal Gaussian process models with data on a grid. The challenge to be addressed is that the spatio-temporal modeling require certain temporal alignment among the data, a property that the wind farm data does not have. Our solution strategy is to construct a shared representative temporal covariate set which not only aligns the temporal inputs but also has a size an order of magnitude smaller than the original data size. With this transformation, our resulting model is able to employ a separable kernel structure that captures both spatial and temporal dependencies. Empirical analysis on a real wind farm dataset shows that our method improves predictive accuracy over existing baselines and can be used to quantify the various impact of the terrain characteristics on turbine performance.
\end{abstract}

\noindent%

\spacingset{1.5}

\vspace{-9 pt}
\section{Introduction}\label{sec:problem_statment}
\vspace{-9 pt}

This paper is concerned with how to build a power curve model for a wind turbine that can incorporate the covariates that characterize the terrain surrounding a wind turbine. Doing so involves addressing new and challenging data science issues.

\begin{figure}[htbp]
    \centering

    % Subfigure 1a
    \begin{subfigure}{0.45\textwidth}
        \centering
        \includegraphics[width=\textwidth]{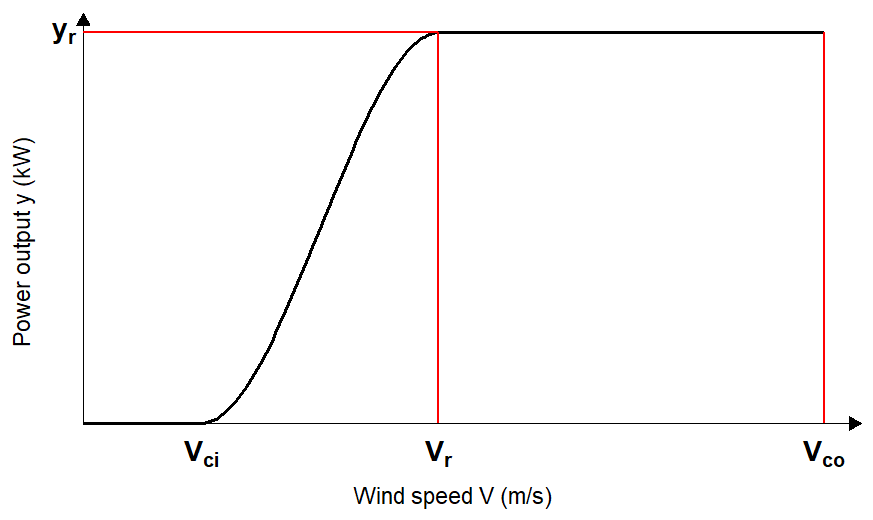}
        \caption{}
        \label{fig:power_curve_nominal}
    \end{subfigure}
    \hfill
    % Subfigure 1b
    \begin{subfigure}{0.54\textwidth}
        \centering
        \includegraphics[width=\textwidth]{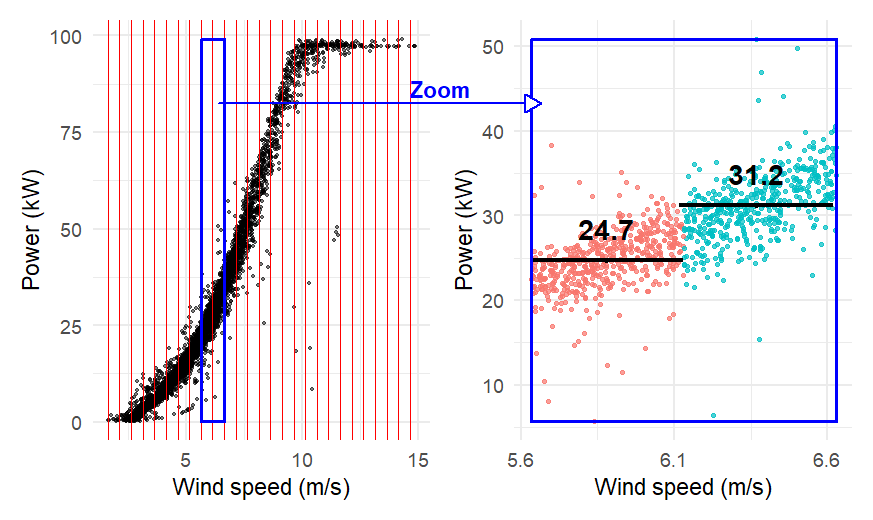}
        \caption{}
        \label{fig:power_curve_binning}
    \end{subfigure}

    \caption{(a) Nominal power curve of a wind turbine \citep{Lee2015a} and (b) IEC binning method to produce a power curve using observed data \citep{Ding2015}}
    \label{fig:power_curves}
\end{figure}

A power curve is a functional curve that links the environmental inputs to a turbine, including the wind speed, to the power generated by the turbine.  Figure~\ref{fig:power_curve_nominal} shows a nominal power curve provided by a wind turbine manufacturer. It is a smooth curve starting at the cut-in wind speed, $V_{ci}$, leveling at the rated wind speed, $V_r$, and ending at the cut-out wind speed, $V_{co}$. The cut-in wind speed is when a turbine starts to produce energy, the rated wind speed is when the turbine's power generation is capped at its designed maximum capacity by using, for instance, pitch control, and the cut-out wind speed is when the turbine is mechanically stopped to avoid structural damage under strong winds. The wind turbine power curve is arguably the most important function used to characterize the production efficiency of a wind turbine or for the purpose of forecasting wind power \citep{Ding2019}.

The turbine manufacturer's \emph{nominal} power curve does not accurately reflect a wind turbine's actual operation and power production. International Electrotechnical Commission (IEC), the international standard body governing the wind industry, recommends estimating power curve using wind and power data collected from an operational turbine~\citep{IEC2005}; the procedure is called the IEC \emph{binning method}.  Figure \ref{fig:power_curve_binning} illustrates the IEC binning method, which cuts wind speed into fine bins, collects the power data in each bin, and uses the sample average of the power data in a bin to represent the power response for that bin. The final power curve is a smooth curve used to connect the bin-wise power averages.  

The essence of the IEC binning method is a one-dimensional nonparametric regression. In the wind industry, wind and power data pairs are collected every 10 minutes, yielding more than 52,000 data points for a year.  With these many data pairs, the binning method produces a reasonably good power curve estimate as long as the input remains univariate, i.e., wind speed alone.  Later research reveals that while wind speed is the dominating factor behind wind power production, other factors matter too, including wind direction, air density, humidity, and turbulence, among others. \cite{Bessa2012} used kernel regression and kernel density estimation to model wind power, considering only a few variables such as wind speed, wind direction, and air density. \cite{Lee2015a} extended this approach with the additive multiplicative kernel (AMK) model, which can accommodate more inputs while maintaining scalability. Chapter 5 of \cite{Ding2019} and its companion \texttt{R} package \citep{DSWE2021} present a comprehensive description of how various machine learning methods can be used for wind turbine power curve modeling as well as how these methods perform on several benchmark datasets.

Nearly all power curve models produced so far---certainly those referenced above---rely on the environmental inputs and power output collected on an operational wind turbine erected at a specific location. One type of factors, i.e., the surface characteristics surrounding a turbine, known as the terrain covariates, are rarely directly incorporated in a power curve model, even though wind engineering researchers \citep{Sempreviva1990, Fragoulis1997, Han2012, Tian2015} seem to all agree that terrain characteristics affect how a wind turbine reacts to inflow wind and thus how it produces power. To the best of our knowledge, the only terrain-incorporating wind power curve model was recently published using a Bayesian hierarchical model \citep[BHM]{Prakash2024}.

Why are there so few terrain-incorporating power curve models, to the extent of near non-existence? Terrain data are different from the rest of the environmental input data mentioned earlier in the sense that they do not change over time, but they are different from turbine to turbine.  In other words, terrain characteristics are time-invariant, spatially-varying-only covariates, whereas the rest of the environmental covariates are both spatially and temporally varying, meaning that they are dynamically changing over time and they could be different from turbine to turbine, too. Mixing the two types of covariates and lumping everything together as inputs in a conventional supervised learning approach does not produce a good terrain-incorporating power curve model. In practice, including terrain features as additional predictors often degrades predictive accuracy. This phenomenon arises because terrain covariates are static, site-specific attributes that are fundamentally different in nature from dynamic environmental inputs, and treating them uniformly as additional predictors in conventional supervised learning frameworks introduces confounding structure that deteriorates predictive performance, as demonstrated empirically in Section S1 of the Supplementary Material.

Admittedly the first to recognize such a unique combination of terrain and non-terrain covariates, \cite{Prakash2024} developed a Bayesian hierarchical power curve model. Their method models a turbine’s power curve using a parametric logistic function with wind speed and temperature as inputs and with two parameters representing the slope and location of the inflection point of the logistic curve, respectively. Their model further expresses these two logistic curve parameters as linear functions of a set of terrain features. Thereby, terrain characteristics are incorporated explicitly. The Bayesian hierarchical power curve model is limited by its reliance on a parametric function form and the linear dependence of curve parameters on terrain features. The Bayesian inference is also computationally demanding, and to meet the practical needs, the current Bayesian hierarchical power curve model uses about 10\% of annual raw data for model inference (yet it still takes more than 2 hours to estimate the model on a typical laptop computer).

These limitations highlight the need for a nonparametric and scalable model that can capture nonlinear relationships between terrain covariates and power production. In this work, we propose a spatio-temporal Gaussian Process (GP) model that explicitly incorporates both time-varying environmental variables and spatial terrain covariates.

Our approach is built upon a set of spatio-temporal models that handle similar combination of spatial and temporal covariates \citep{banerjee2008, Cressie2011, heaton2019}.  As the detailed literature review in Section~\ref{sec:literature} will show, the existing spatio-temporal models require some kind of temporal alignment among their data, which is missing in the wind turbine application. That turns to be the major challenge we have to overcome. We address this challenge by applying a pseudo input and output approach, which allows us to align turbine observations along the temporal dimension with a much reduced data size. Once the pseudo inputs and outputs are constructed, a separable spatio-temporal model in the fashion of \citet{banerjee2008, Cressie2011, heaton2019} can be established and used as the terrain-incorporating power curve model.

The remainder of the paper is structured as follows. Section~\ref{sec:literature} introduces the data and reviews relevant literature. Section~\ref{sec:model} presents the formulation and solution for a spatio-temporal GP model conditioned on a set of pseudo inputs and outputs Section~\ref{sec:experiments} reports experiments evaluating performance and the benefits of the terrain-aware model. Section~\ref{sec:conclusion} concludes with directions for future work.

\section{Data and Literature Review}\label{sec:literature}
\vspace{-9 pt}
The dataset used in this study originates from an onshore wind farm comprising 66 turbines, the same wind farm as described in \citet{Prakash2024}. Figure 2 in \citet{Prakash2024} presents the layout of the turbine on that wind fram. 

The dataset includes the 10-minute recordings of power, wind speed, wind direction, temperature, turbulence intensity, and wind shear for the years of 2017 and 2018. For each turbine, approximately 45,000 records are available per year. This is fewer than the 52,560 possible 10-minute intervals per year because there are missing data and also because the intervals during which the turbines were non-operational have been removed. 

Let each turbine be indexed by \( i \in \{1, \dots, m\} \) and $m=66$ for this wind farm. For turbine \( i \) at time $t$, the wind and environmental inputs, as mentioned above, are denoted by a $p \times 1$ vector $\mathbf x^{(i)}_t = (x^{(i)}_{1,t}, \ldots, x^{(i)}_{p,t})^\top$.  Suppose that there are $n_i$ temporal observations for turbine \( i \), i.e., $t=1, \ldots, n_i$, then we can aggregate all temporal inputs in a matrix \(\mathbf{X}^{(i)} \in \mathbb{R}^{p \times n_i}\), such that
\begin{equation}
\mathbf X^{(i)} =\begin{pmatrix} | & | & \cdots & | \\ \mathbf x^{(i)}_1 & \mathbf x^{(i)}_2 & \cdots & \mathbf x^{(i)}_{n_i} \\ | & | & \cdots & | \end{pmatrix}.
\end{equation}
The corresponding power output is represented by $y_t^{(i)}$ or aggregated as $\mathbf y^{(i)} \in \mathbb{R}^{n_i}$. Both \(\mathbf{x}^{(i)}_t\) and \(y^{(i)}_t\) vary over time and from turbine to turbine.

In addition to $\{\mathbf x^{(i)}_t, y^{(i)}_t\}$ pairs, terrain characteristics around each turbine are available for this wind farm. IEC identifies three key terrain descriptors~\citep{IEC2013}: slope (in degrees), ruggedness index (RIX), which is a dimensionless measure of terrain roughness, and the maximum ridge height (in meters). These terrain descriptors are computed for 36 directional sectors with a 10° increment; please refer to Section~S2 in the Supplementary Material for a graphical illustration of the sectional partition around a turbine. At the finest level, the terrain surrounding a turbine is characterized by a $36 \times 3$ matrix, where each row corresponds to a wind direction sector and each column to one terrain descriptor.  At the next level, the characterization is averaged over all wind direction sections, producing a $3\times 1$ descriptor vector, $\mathbf s_i$.  Further up, a terrain is classified into five roughness categories, with 1 being most flat and 5 being most complex. \citet[Section VI-C4]{IEC2013} provides specific rules of how a terrain is classified based on numerical measurements.

Instead of using the simple average over the 36 wind directions, \citet{Prakash2024} used a wind-direction-weighted average to produce the vectorial representation of the three terrain descriptors for a turbine. Let $\mathbf s_{b,i}$ be the $3 \times 1$ vector of terrain characteristics for direction sector $b$ at turbine $i$ and $\gamma_{b,i}$ be the empirical frequency of wind direction in sector $b$ at turbine $i$. Then the direction-weighted terrain description vector is $\mathbf s_i = \sum_{b=1}^{36} \gamma_{b,i} s_{b,i}$. In this study, we use the same weighted description of terrain as in \citet{Prakash2024}. 

Having described both the temporal and spatial components of the data, the objective for terrain-incorporating power curve modeling is to find a nonparameteric function $f^{(i)}(\cdot)$ for turbine $i$, such that
\begin{equation}\label{equation2}
y_t^{(i)} = f^{(i)}(\mathbf x_t^{(i)}, \mathbf s_i) + \epsilon_t^{(i)}, \quad i = 1, \ldots, m, \,\, \text{and} \,\, t= 1, \ldots, n_i,
\end{equation}
where $\mathbf x_t^{(i)}$ are both spatially and temporally varying but $\mathbf s_i$ is spatially varying and temporally invariant. Some may question why $\mathbf s_i$ is temporally invariant if vegetation around the turbine changes seasonally.  The answer is that $\mathbf s_i$ measures ground surface changes much larger in scale than vegetation's seasonal change.  Recall that a commercial wind turbine has a typical hub height of 80 meters, whereas seasonal vegetation is within 1--2 meters of the ground, so that the vegetation's seasonal change does not practically affect $\mathbf s_i$. 

Spatio-temporal Gaussian process (STGP) models provide a promising framework for capturing both temporal and spatial dependencies \citep{banerjee2008, Cressie2011, heaton2019}. Inclusion of both spatial and temporal covariates leads to a large covariance matrix, creating modeling and computational difficulties. A common approach to enable tractable inference for large datasets is to adopt a separable kernel structure:
\begin{equation}
\mathbf{R} = \mathbf{R}_x \otimes \mathbf{R}_s,
\end{equation}
where \( \mathbf{R}_s \in \mathbb{R}^{m \times m} \) models spatial similarity across \( m \) locations and \( \mathbf{R}_x \in \mathbb{R}^{n \times n} \) models correlations across \( n \) temporal time points. This Kronecker formulation, rooted in early work on separable space-time covariance structures \citep{stein2005, genton2007} and later applied to GPs for scalability \citep{Saatci2011,Luttinen2012}, dramatically reduces the cost of matrix operations and enables efficient computation.

However, a key assumption underlying the Kronecker-based STGP models is that observations exist for every pair \( ( \mathbf{x}, \mathbf{s}) \), so the data forms a complete, fully observed spatio-temporal grid. In other words, for any timestamp, $y$ is observed at all spatial locations, whereas for any spatial location, $y$ is also observed for all timestamps. Please see Figure \ref{fig2:a} for an illustration. But the wind turbine spatio-temporal data is not on a fully observed grid. If arranged on a grid, it would have blocks of missing data; see Figure \ref{fig2:b}.  
\begin{figure}[ht]
\centering
\small

% ----------- Subfigure (a) on top -----------
\begin{subfigure}[t]{0.6\textwidth}
\centering
{\renewcommand{\arraystretch}{1.3}
\setlength{\tabcolsep}{4pt}
\scriptsize
\begin{tabular}{c|cccc}
    & $\mathbf{x}_1^{(1)}$ & $\mathbf{x}_2^{(2)}$ & $\cdots$ & $\mathbf{x}_{n_1}^{(m)}$ \\
    \midrule
    $\mathbf{s}_1$ & $y_1^{(1)}$ & $y_2^{(1)}$ & $\cdots$ & $y_{n_1}^{(1)}$ \\
    $\mathbf{s}_2$ & $y_1^{(2)}$ & $y_2^{(2)}$ & $\cdots$ & $y_{n_1}^{(2)}$ \\
    $\vdots$       & $\vdots$ & $\vdots$ &         & $\vdots$ \\
    $\mathbf{s}_m$ & $y_1^{(m)}$ & $y_2^{(m)}$ & $\cdots$ & $y_{n_1}^{(m)}$ \\
    \bottomrule
\end{tabular}
}
\caption{The ideal fully observed grid.}
\label{fig2:a}
\end{subfigure}

\vspace{1em} % space between top and bottom

% ----------- Subfigure (b) on bottom -----------
\begin{subfigure}[t]{0.9\textwidth}
\centering
{\renewcommand{\arraystretch}{1.3}
\setlength{\tabcolsep}{3pt}
\scriptsize
\begin{tabular}{c|cccc|cccc|c|cccc}
    & \multicolumn{4}{c|}{\textbf{Turbine} 1}
    & \multicolumn{4}{c|}{\textbf{Turbine} 2}
    & $\cdots$
    & \multicolumn{4}{c}{\textbf{Turbine} $m$} \\
    \toprule
    & $\mathbf{x}_1^{(1)}$ & $\mathbf{x}_2^{(1)}$ & $\cdots$ & $\mathbf{x}_{n_1}^{(1)}$
    & $\mathbf{x}_1^{(2)}$ & $\mathbf{x}_2^{(2)}$ & $\cdots$ & $\mathbf{x}_{n_2}^{(2)}$
    & $\cdots$
    & $\mathbf{x}_1^{(m)}$ & $\mathbf{x}_2^{(m)}$ & $\cdots$ & $\mathbf{x}_{n_m}^{(m)}$ \\
    \midrule
    $\mathbf{s}_1$
    & $y_1^{(1)}$ & $y_2^{(1)}$ & $\cdots$ & $y_{n_1}^{(1)}$
    & NA & NA & $\cdots$ & NA
    & $\cdots$
    & NA & NA & $\cdots$ & NA \\
    $\mathbf{s}_2$
    & NA & NA & $\cdots$ & NA
    & $y_1^{(2)}$ & $y_2^{(2)}$ & $\cdots$ & $y_{n_2}^{(2)}$
    & $\cdots$
    & NA & NA & $\cdots$ & NA \\
    $\vdots$
    & $\vdots$ & $\vdots$ &         & $\vdots$
    & $\vdots$ & $\vdots$ &         & $\vdots$
    & $\cdots$
    & $\vdots$ & $\vdots$ &         & $\vdots$ \\
    $\mathbf{s}_m$
    & NA & NA & $\cdots$ & NA
    & NA & NA & $\cdots$ & NA
    & $\cdots$
    & $y_1^{(m)}$ & $y_2^{(m)}$ & $\cdots$ & $y_{n_m}^{(m)}$ \\
    \bottomrule
\end{tabular}
}
\caption{Structured missingness in wind turbine data. Each turbine observes its own set of temporal inputs \( \mathbf{x} \) and outputs $y$, while other turbine blocks remain unobserved (NA).}
\label{fig2:b}
\end{subfigure}

\caption{Illustration of structured missingness: (a) the ideal fully observed grid and (b) the actual observed blocks with missing entries for turbine data.}
\label{fig2}
\end{figure}

Research developments have been reported for handling missing data over a spatio-temporal grid for the purpose of STGP modeling. We refer to this branch of research as \emph{STGP on a partially observed, incomplete grid}.  Our literature review shows that STGP on a partially observed, incomplete grid can be grouped into the following schools of thoughts: 
\begin{itemize}
\item \textbf{Simple imputation}. \citet{lambardi2022} impute missing sensor readings using the Last Observation Carried Forward (LOCF) method, which fills each missing value with the most recent observed value. This allows them to maintain a fully observed grid for Kronecker-based GP inference. 
\item \textbf{Bayesian filling of missing data}. \citet{gu2022} introduces a Bayesian latent factor model where each observation matrix is factorized into spatial and temporal components, and missing values are inferred jointly via GP priors on latent dynamics. 
\item \textbf{Induction and interpolation}. \citet[KISS-GP]{wilson2014} and \citet{wilson2015} approximate the full covariance matrix using interpolation from a grid of inducing points. The covariance matrix is approximated as \( R \approx W R_{UU} W^\top \), where \( W \) is a sparse interpolation matrix and \( R_{UU} \) is the covariance between inducing points. Missing values are not imputed directly; instead, the GP smoothness prior is used to locally interpolate through spatially nearby points. The Fixed Rank Kriging (FRK) framework is a popular method in spatial statistics \citep{cressie2008}. In later developments, FRK has been extended to the spatio-temporal domain and is implemented through discretization into basic areal units (BAUs), with basis functions evaluated at their centroids \citep{zammitmangion2021}. Predictions are obtained by interpolating through the smooth surface implied by these basis functions.
\item \textbf{Filtering and smoothing}. \citet{todescato2020} proposed, and \citet{hamelijnck2021} extended, a Kalman filtering-based approach, which reformulates GP inference in temporal domains using state-space models. This is possible because GPs with Matérn kernels are equivalent to solutions of linear stochastic differential equations. The GP is thus represented as a dynamical system, and inference can be performed recursively using Kalman filtering and smoothing, reducing the cost from \( \mathcal{O}(n^3) \) to \( \mathcal{O}(n)\). Missing temporal data is naturally handled in this framework---at missing time steps, the Kalman update step is skipped and the prior prediction is simply propagated forward. 
\item \textbf{Mask-out projection} \citet{lin2024} modifies Kronecker-based GP inference by introducing a diagonal projection matrix \( P \). This matrix masks out unobserved entries from the likelihood and gradient calculations. The Kronecker structure of the kernel \( R = R_t \otimes R_s \) is retained, but inference is restricted to observed entries via \( P^\top R P \). This approach avoids imputing or marginalizing missing values and maintains computational scalability.
\end{itemize}

Unfortunately, the existing methods are ill-equipped to handle the wind turbine spatio-temporal data due to the uniqueness of the problem.  The first is the \emph{structured missingness}. If one looks at Figure \ref{fig2:b}, it is easy to notice blocks of data are missing, instead of random or scattered gaps over the grid. The blocks of missing data are due to different temporal variates being used in the wind turbine spatio-temporal model.  

Recall that classical STGP models define the temporal input as time indices and the spatial input as geographic coordinates (latitude and longitude). But in wind energy applications, power generation is influenced by environmental covariates rather than timestamps. The temporal input in Equation~(\ref{equation2}), $\mathbf x_t^{(i)}$, is the vector that includes wind speed, temperature, and other covariates. Unlike timestamps, $\mathbf x_t^{(i)}$ is different across the turbines even at the same time $t$. What this suggests is that each turbine effectively observes a distinct temporal input at each time, and as a result, a common temporal input \( \mathbf{x}^* \) is not shared across turbines. The number of observations per turbine, $n_i$, also varies across turbines, due to different periods of non-operation, outage, and/or when sensors malfunction. 

Such structured missing of data causes substantial problems when one attempts to use the existing methods to handle the wind data.  For example, the simple imputation method, LOCF, works well when missing data are random and fewer than 10\% of the total amount. The wind data, by comparison, is 98.5\% missing when put on a spatio-temporal grid.

The induction and interpolation methods, like KISS-GP and the Kalman-based GPs, work well for sparse or scattered data gaps on a grid, but they break down when large structured blocks are missing or when the inputs are not well aligned with the grid, as in the case of wind data. As for methods like FRK, its model must still estimate and tune basis weights for BAUs.  But when there are little or no nearby observations because of blocks of missing data, such estimation and tuning are of poor quality and unreliable. On top of that, FRK takes longitude and latitude as spatial inputs and time as the temporal input, i.e., two spatial dimensions and one temporal dimension; such model setup makes it inapplicable to the wind farm data of which the spatial dimension (terrain) is three and the temporal dimension (the number of covariates in $\mathbf x$) could be much more than one.

For the masked-out projection approach, if the overlap between the spatial and temporal components becomes too sparse---like too few turbine-time combinations are jointly observed---the projection matrix \( P \) becomes too small or even empty, and then the projection method will fail. That indeed happened---when we tried the masked-out projection on the wind data, the projected Kronecker GP collapsed.

The Bayesian filling of missing data can, in principle, allow structured missingness to be handled, but it relies on expensive MCMC and lacks scalability.  For all work reviewed above, their $m\times n$ ranges from a few hundred \citep{lin2024} to 115,000 \citep{gu2022}. For the wind application at hand, $m$ is 66 and $n$ is about 45,000, so that $m\times n$ is of the order of 3 million.  No existing STGP method can handle this scale.

\vspace{-9 pt}
\section{Methodology}\label{sec:model}
\vspace{-9 pt}

An idea to address the wind farm data is as follows.  Suppose that we could have a set of data $\{\mathbf{x}_t^*, \mathbf s_i, y_t^{i*}\}$, for $t=1, \ldots, n_*$ and $i=1, \ldots, m$, where $\mathbf{x}_t^*$ is a set of environmental inputs common to all turbines and $y_t^{i*}:=y^{(i)}(\mathbf x^*_t, \mathbf s_i)$ is the corresponding power output at time $t$ and turbine $i$.  Such data can be arranged onto a fully observed grid just as that in Figure~\ref{fig2:a}; see Table ~\ref{fig:complete_table}, and then, the existing separable STGP can be applied.

\begin{table}[ht]
\centering
\renewcommand{\arraystretch}{1.4}
\setlength{\tabcolsep}{5pt}
\small
\caption{With $\{\mathbf{x}_t^*, \mathbf s_i, y_t^{i*}\}_{t=1; \,\, i=1}^{n_*; \,\,\,\,\, m}$, the wind data forms a complete grid.}
\vspace{10 pt}
\begin{tabular}{c|cccc}

    & $\mathbf{x}_1^*$ & $\mathbf{x}_2^*$ & $\cdots$ & $\mathbf{x}_{n_*}^*$ 
    \\
    \midrule
    $\mathbf{s}_1$ 
    & $y_1^{1*}$ & $y_2^{1*}$ & $\cdots$ & $y_{n_*}^{1*}$ \\

    $\mathbf{s}_2$ 
    & $y_1^{2*}$ & $y_2^{2*}$ & $\cdots$ & $y_{n_*}^{2*}$ \\

    $\vdots$ 
    & $\vdots$ & $\vdots$ &        & $\vdots$ \\

    $\mathbf{s}_m$ 
    & $y_1^{m*}$ & $y_2^{m*}$ & $\cdots$ & $y_{n_*}^{m*}$ \\

    \bottomrule
\end{tabular} \label{fig:complete_table}
\end{table}

There is a requirement and a hurdle. The requirement for computational feasibility is that $n_*$ needs to be much smaller than $n_i$.  Recall that for the wind application at hand, $m\times n_i$ is of the order of 3 million, which is too large for an STGP to be computationally feasible. Practically, $n_*$ needs to be at least an order of magnitude smaller than $n_i$.

The bigger hurdle is how to get $\{\mathbf{x}_t^*, y_t^{i*}\}$. The requirement on a small $n_*$ points to sampling a subset. But the subset approach does not work because we must have a common set of $\mathbf{x}_t^*$ to avoid the structured missingness. Since we have 66 turbines, it is practically impossible to obtain a decent common set of $\mathbf x$'s. What is possible is to create a pseudo input set, which could happen in the physical world but does not exist in the observed dataset.  The $y_t^{i*}$, the corresponding response of turbine $i$ under $\mathbf{x}_t^*$, also does not exist in the observed dataset and is thus a pseudo output.  

Let us denote the pseudo inputs collectively as $\mathbf x^* = \{\mathbf x_t^*\}$ and the pseudo outputs collectively as $\mathbf y^* = \{y_t^{i*} \}$ for $t=1, \ldots, n_*$ and $i=1, \ldots, m$. Ideally, one could optimize a loss function, $\mathcal{L}(\cdot, \cdot)$, and simultaneously solve the problem of choosing the pseudo inputs and pseudo outputs, as well as estimating the parameters for the STGP model, denoted by $\theta$, such as
\begin{equation}
\min_{\mathbf{x}^*, \, \mathbf y^*, \, \theta} \quad \frac{1}{\sum_{i=1}^m n_i}\sum_{i=1}^m \sum_{t=1}^{n_i} \mathcal{L}\Big(y_t^{(i)} , f^{(i)}(\mathbf x_t^{(i)}, \mathbf s_i|\mathbf{x}^*, \mathbf y^*, \boldsymbol{\theta})\Big).
\end{equation} 
We find that jointly optimizing the pseudo inputs/outputs and the covariance parameters is extremely challenging. In this paper, we instead follow a divide-and-conquer strategy, that is, we first select $\{\mathbf{x}_t^*\}$  using a proxy criterion, then estimate $\{y_t^{i*}\}$, and finally, solve for $\boldsymbol{\theta}$ conditioned on $\{\mathbf{x}_t^*\}$ and $\{y_t^{i*}\}$ chosen in the early steps. We acknowledge that our approach does not guarantee the optimal solution, but our numerical evidence shows that such a sequential procedure produces a working terrain-incorporating wind power curve with appreciable improvement over alternative methods. The next three subsections present the details for each of the actions.

We note that the concept of using pseudo inputs and outputs is not new.  In the literature of GP approximation for fast computation, one school of thought is to use the pseudo-inputs, also known as inducing points. Early examples include \cite{snelson2005,snelson2007} and \cite{titsias2009}, where the inducing locations are optimized jointly with covariance hyperparameters or updated through greedy or EM procedures. Recent methods such as \cite{Liu2025} follow a similar iterative structure by repeatedly updating parameters and computing weights for all observations. These pseudo-input GP approximations still scale directly with the size of the full data. A joint optimization is doable in these approaches because their full data is in a much smaller amount (like, tens of thousands). Our application, by contrast, has millions of observations, and as a result, a joint optimization is prohibitively expensive. The divide-and-conquer strategy we will employ instead accesses the full data only once for determining the pseudo-input and pseudo-output sets.  In both steps, the computational complexity depends only on the number of pseudo-locations, not on the full data size. As a result, all subsequent computations scale solely with the chosen pseudo-set size, making the resulting method practical for large datasets.

%Our solution strategy focuses on the two primary issues: the structured data missingness and a large covariance matrix. To address these, our main idea to construct a shared representative temporal covariate set, called $\mathbf x_t^*$, which not only aligns the temporal inputs but also has a size an order of magnitude smaller than the original $n_i$. This construction serves as one stone killing two birds. Subsequently, a STGP is built using this newly constructed temporal set, which is our terrain-incorporating power curve model.

\subsection{Decide the Pseudo Input Set}
\vspace{-9 pt}

The first step in our approach is to find the common pseudo input set $\mathbf x^*_t$ of size $n_* \times 1$, for $t=1, \ldots, n_*$, which will be used to replace $\mathbf x_t^{(i)}$ for $t=1,\ldots,n_i$ and $i=1, \ldots, m$. Let us use $\mathbf X^*$ as the aggregation of $\mathbf x^*_t$'s, such that
\begin{equation}
\mathbf X^* =\begin{pmatrix} | & | & \cdots & | \\ \mathbf x^*_1 & \mathbf x^*_2 & \cdots & \mathbf x^*_{n_*} \\ | & | & \cdots & | \end{pmatrix}_{p\times n_*}.
\end{equation}
Our mission to construct the common pseudo input set is to find the following mapping:
$$\left\{ \mathbf X^{(i)}\right\}_{i=1}^m \rightarrow \mathbf X^*.$$

For this purpose, we adopt the \textit{support points} \citep{Mak2018}, which offers a statistically principled approach to representative data selection. Support points provide a set of samples that optimally represent a continuous distribution. In our context, this yields a compressed yet distributionally faithful approximation of the union of all \( \mathbf{x}^{(i)}_t \), preserving key temporal characteristics while significantly reducing the temporal input size.

Support points can be obtained by minimizing the energy distance \citep{szekely2013} between the empirical distribution $F$ of the pooled inputs \begin{equation}\label{eq:pooled-set}
\mathcal{X} \;=\; \bigcup_{i=1}^m \{\mathbf{x}^{(i)}_1,\dots,\mathbf{x}^{(i)}_{n_i}\},
\end{equation}
and the empirical distribution $F_{n_*}$ of the representative set:
\begin{equation}\label{eq:emp-energy}
\widehat{E}_N(F,F_{n_*})
\;=\;
\frac{2}{N n_*}\sum_{r=1}^{N}\sum_{k=1}^{n_*}\!\big\|\mathbf{x}_r-\mathbf{x}^*_k\big\|
\;-\;\frac{1}{N^2}\sum_{r=1}^{N}\sum_{r'=1}^{N}\!\big\|\mathbf{x}_r-\mathbf{x}_{r'}\big\|
\;-\;\frac{1}{n_*^2}\sum_{k=1}^{n_*}\sum_{l=1}^{n_*}\!\big\|\mathbf{x}^*_k-\mathbf{x}^*_l\big\| ,
\end{equation}
where  $\|\cdot\|$ is the Euclidean norm, $N \;=\; \sum_{i=1}^m n_i$, $\{\mathbf{x}_r\}_{r=1}^{N} \subset \mathcal{X}$, and $\mathbf{x}^*_1,\dots,\mathbf{x}^*_{n_*}$ are the support points to be learned. The first term concentrates $\mathbf{X}^*$ in high-density regions of $F$, the third (negative) term spreads the points to avoid clustering, and the middle term depends only on $\{\mathbf{x}_r\}$ (constant with respect to $\{\mathbf{x}^*_k\}$). \citet{Mak2018} minimize \eqref{eq:emp-energy} via a majorization–minimization (MM) algorithm where each iteration constructs a tight surrogate of $\widehat{E}_N(F,F_{n_*})$ and yields closed-form coordinate updates for the support points, guaranteeing monotone decrease of the objective. This provides practical runtimes for moderate $n_*$ even when $N$ is large.

The number of support points, $n_*$, is tunable; smaller values improve scalability but risk information loss, whereas larger values preserve more detail but increase computational burden. We set $n_* = 1192$, determined via a standard elbow rule to balance predictive accuracy and computational efficiency, as described in the Supplementary Material (S3).

\vspace{-9 pt}
\subsection{Estimate the Pseudo Output Set}
\vspace{-9 pt}

Once the common pseudo inputs, \( \mathbf{x}_t^* \), are constructed, a subsequent question is what response would be corresponding to these \( \mathbf{x}_t^* \).  We decide to approximate $ y^{(i)}(\mathbf x^*_t)$ using a turbine-specific, temporal only power curve. 

Recall that there are mature machine learning methods for modeling turbine-specific, temporal only power curves; see Chapter 5 of \cite{Ding2019}. The process is to take the training data pairs $\{\mathbf x_t^{(i)}, y_t^{(i)}\}_{t=1}^{n_i}$ and run a nonlinear regression for $i=1,\ldots,m$.  The resulting regression function $\hat{f}_\text{temp}^{(i)}(\cdot)$ is a special form of the spatio-temporal function $f^{(i)}(\cdot, \cdot)$ in Equation~(\ref{equation2}).  Once $\hat{f}_\text{temp}^{(i)}(\cdot)$ is trained, then \( y^{(i)}(\mathbf x^*_t) \) can be approximated by
$$ y^{i*}_t:=\hat{y}^{(i)}(\mathbf x^*_t) = \hat{f}_\text{temp}^{(i)}(\mathbf x^*_t),\; i=1,\ldots,m; \; t=1,\ldots,n_*.$$

Among the many options for modeling the temporal-only power curve, we choose \emph{twinGP} \citep{Vakayil2023}. TwinGP is a scalable Gaussian process approximation that produces competitive predictive accuracy as measured by out-of-sample RMSE on held-out test sets. The reason for our choice is that twinGP is the fastest method among the set of methods that show superior predictive performance.  The fast computation is highly desirable when handling a wind farm of 66 wind turbines.  

\cite{Prakash2023} raised an issue of regression on autocorrelation data, i.e., the issue of temporal overfitting. This is to say, when the data is autocorrelated, the regression term may not remove all autocorrelation in residuals, leaving the error term still autocorrelated and thus violating the standard model assumption in regression problems.  As a result, using the standard cross-validation or alike tends to choose the wrong model.  

To better handle autocorrelated data and avoid temporal overfitting, \cite{chokhachian2026} offered the \emph{thinned twinGP} model. They partition the training sequence into $B$ blocks following the data binning idea used in \cite{Prakash2023}. Within each block, the data points are sufficiently separated in sequence so that the within-block data autocorrelation is weakened to the extent that one can assume the absence of correlation within each block.  For each test input, they fit a twinGP model on every block and then average the predictions from the $B$ resulting twinGP models. This revision is simple enough to carry out and still enjoys fast computation. Training this revised twinGP for all $B$ blocks takes roughly 1.5 minutes per turbine on an annual dataset of size 45,000 data pairs, or about one and a half hours for all 66 turbines on this wind farm, making it well-suited for handling the large-scale wind turbine datasets. 

The twinGP provides both a predictive mean and a predictive variance at each pseudo-observation. The predictive variance of the thinned twinGP, $v_t^{(i)}$, comprises variance contribution from both GP process variability and observation noise. To better handle the uncertainty in the second-stage spatio-temporal model, we decompose this predictive variance into two components: the heteroscedastic observation noise and the latent-process variance components. 

\vspace{9 pt} 

\begin{algorithm}[H]
\caption{10-NN residual estimator of $\eta_t^{(i)}$ for turbine $i$}
\label{alg:10nn_sigma2_notrans}
\small
\KwIn{
Training data $\{(\mathbf x_u^{(i)},y_u^{(i)})\}_{u=1}^{n_i}$;
pseudo inputs $\{\mathbf x_t^*\}_{t=1}^{n_*}$; neighborhood size $k=10$.
}
\KwOut{
Estimated observation-noise variances $\{\hat\eta_t^{(i)}\}_{t=1}^{n_*}$.
}

\textbf{Step 0 (fit temporal model)}  
Fit the thinned twinGP on $\{(\mathbf x_u^{(i)}, y_u^{(i)})\}_{u=1}^{n_i}$ and denote the predictive mean by $\hat{\mu}^{(i)}(\cdot)$.\;

\textbf{Step 1 (10-NN neighborhoods)}  
For each $t$, compute $N_t^{(i)}$ as the $k$ nearest neighbors of $\mathbf x_t^*$ using Euclidean distance.\;

\textbf{Step 2 (residuals)}  
For $u \in U^{(i)}=\cup_t N_t^{(i)}$,
\[
\hat{y}_u^{(i)} = \hat{\mu}^{(i)}(\mathbf x_u^{(i)}), \qquad
r_u^{(i)} = y_u^{(i)}-\hat{y}_u^{(i)}.
\]

\textbf{Step 3 (local variance)}  
For each $t$,
\[
\hat\eta_t^{(i)} =
\operatorname{Var}\!\left(\{r_u^{(i)}:\ u\in N_t^{(i)}\}\right).
\]

\Return $\{\hat\eta_t^{(i)}\}_{t=1}^{n_*}$.
\end{algorithm}

\vspace{9 pt}

We approximate the heteroskedastic observation-noise variance, denoted by $\eta_t^{(i)}$, using a $k$-nearest neighbor-based algorithm; see Algorithm~\ref{alg:10nn_sigma2_notrans}.  Then, the pseudo-output latent-process variance, $\eta_{*,t}^{(i)}$, is estimated as
\begin{equation}
\eta_{*,t}^{(i)} = v_t^{(i)} - \eta_t^{(i)}.
\end{equation}

\begin{figure}[ht]
\centering
\includegraphics[width=.6\textwidth]{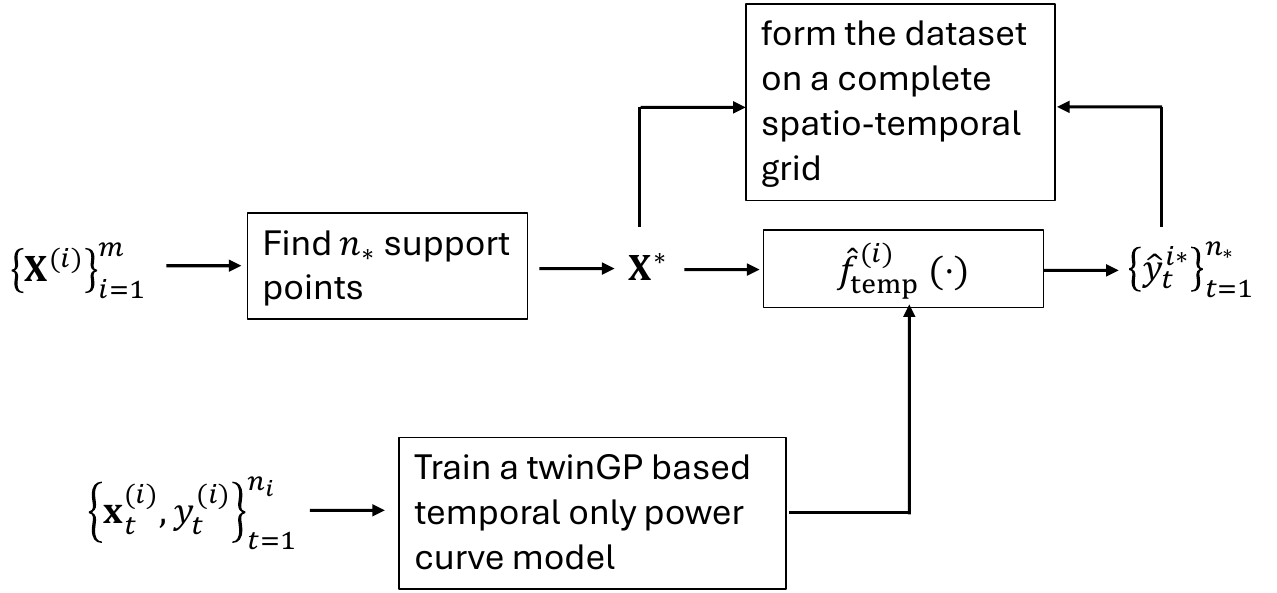}
\caption{Constructing the pseudo inputs and outputs.}\label{fig:sup+TwinGP}
\end{figure}

Once the pseudo outputs, corresponding to the pseudo inputs, are constructed, the wind spatio-temporal data can be arranged on a complete, fully observed grid, just as shown in Table \ref{fig:complete_table}. The overall procedure for constructing the pseudo inputs/outputs is explained in Figure \ref{fig:sup+TwinGP}. This procedure of data transformation offers an additional benefit. In the original dataset, the total variance in the power output data can be attributed to four sources:  
\begin{enumerate}
\item temporal variation over time within each turbine,  
\item variation in temporal inputs across turbines,  
\item variation in static spatial features (e.g., terrain), and  
\item observational noise.
\end{enumerate}
After using the pseudo inputs/outputs, we effectively eliminate the inter-turbine variation in temporal inputs. This simplification reduces input space heterogeneity and allows the model to focus more effectively on the remaining sources of variation, specifically spatial heterogeneity and intra-turbine temporal variation. 

\vspace{-18 pt}
\subsection{Spatio-Temporal GP Model}
\vspace{-9 pt}

With the power outputs from all turbines aligned on a common temporal grid and paired with their respective spatial (terrain) features, we are now equipped to construct a separable STGP conditioned on the pseudo inputs and outputs. Here we add a note clarifying the notation  in Eq.~\eqref{equation2} in which the original purpose is to fit a turbine-specific power curve function, \(f^{(i)}(\cdot)\). Because learning the effect of \(\mathbf{s}\) requires the use of multiple turbines associated with different terrains, the resulting power curve function, \(f(\cdot)\), is learned for the whole farm, suggesting that the superscript, $(i)$, can be dropped, and the turbine-specific features enter the resulting power curve function through its inputs $\mathbf x_t^{(i)}$ and $\mathbf s_i$.

Let \( \mathbf{Y}^* \in \mathbb{R}^{m \times n_*} \) represent the aligned power outputs over \( m \) turbines and \( n_* \) temporal inputs. That is, $\mathbf{Y}^*$ is the matrix containing those $\hat{y}$'s in Table~\ref{fig:complete_table}. We define $\operatorname{vec}(\mathbf{Y}^*)$ as the column-wise stacking of $\mathbf{Y}^*$, obtained by stacking its columns sequentially into a long column in $\mathbb{R}^{mn_*\times1}$. Let \( \mathbf{X}^* \in \mathbb{R}^{ p \times n_*} \) be the temporal covariates (e.g., wind speed, temperature), and \( \mathbf{S} \in \mathbb{R}^{m \times q} \) be the spatial covariates (e.g., terrain features). We assume the following matrix-variate GP model \citep{Bonilla2007, Rakitsch2013}, which corresponds to an ordinary kriging (OK) model with separable covariance \citep{Cressie1991}:
\begin{equation}
\mathrm{vec}(\mathbf{Y}^* ) \sim \mathcal{N}_{mn_*}(\mu \mathbf{1}_{mn_*}, \sigma^2 \mathbf{R}_x \otimes \mathbf{R}_s),
\end{equation}
where, \( \mu \in \mathbb{R} \) denotes a constant mean, and \( \sigma^2 \) represents the process variance. 
The temporal dependence is captured by the kernel matrix \( \mathbf{R}_x \in \mathbb{R}^{n_* \times n_*} \) defined over \(\mathbf{X}^*\), 
while the spatial dependence is described by the kernel matrix \( \mathbf{R}_s \in \mathbb{R}^{m \times m} \) defined over \(\mathbf{S}\). 
The operator \( \otimes \) denotes the Kronecker product.

In this work, we choose the temporal kernel matrix $\mathbf{R}_x \in \mathbb{R}^{n_* \times n_*}$ 
based on a product exponential kernel:
\begin{equation}
[\mathbf{R}_x]_{uv} = \exp\!\left(-\sum_{l=1}^p \frac{\lvert x^*_{lu} - x^*_{lv} \rvert}{\theta_{x,l}}\right),
\end{equation}
where $x^*_{lu}$ denotes the $l$-th temporal covariate of the $u$-th input. The entries of the spatial kernel matrix $\mathbf{R}_s \in \mathbb{R}^{m \times m}$ 
are defined using a rational quadratic form \citep{Rasmussen2006}:

\begin{equation}
[\mathbf{R}_s]_{ab} =
\left[ 1 + \frac{1}{\alpha}\sum_{g=1}^q 
\left( \frac{s_{ag}-s_{bg}}{\theta_{s,g}} \right)^2 
\right]^{-\alpha},
\end{equation}
where $s_{ag}$ denotes the $g$-th spatial covariate of the turbine $a$. We fix $\alpha=1$. The reason for these specific choices of covariance functions is provided in Supplementary Material (S4).

\vspace{-9 pt}
\subsubsection{Parameter Estimation}
\vspace{-9 pt}

Let \( \mathbf{R}=\mathbf{R}_x\otimes\mathbf{R}_s \). Using the properties of Kronecker products \citep{Petersen2012},
%Using Kronecker identities \citep{Petersen2012}:
%\begin{equation}
%\begin{aligned}
%(\mathbf{R}_x^{-1} \otimes \mathbf{R}_s^{-1})\,\mathrm{vec}(\mathbf{Y}^*)
%&= %\mathrm{vec}\!\big(\mathbf{R}_s^{-1}\,\mathbf{Y}^*\,\mathbf{R}_x^{-%1}\big),
%\end{aligned}
%\end{equation}
we can efficiently estimate the GP model parameters as:

\begin{equation}
\begin{split}
\hat{\mu}
&= \frac{\mathbf{1}_{m n_*}^\top \mathbf{R}^{-1}\,\mathrm{vec}(\mathbf{Y}^*)}
        {\mathbf{1}_{m n_*}^\top \mathbf{R}^{-1}\,\mathrm{vec}(\mathbf{1}_m \mathbf{1}_{n_*}^\top)} \\
&= \frac{\mathbf{1}_{m n_*}^\top (\mathbf{R}_x^{-1} \otimes \mathbf{R}_s^{-1})\,\mathrm{vec}(\mathbf{Y}^*)}
        {\mathbf{1}_{m n_*}^\top (\mathbf{R}_x^{-1} \otimes \mathbf{R}_s^{-1})\,\mathrm{vec}(\mathbf{1}_m \mathbf{1}_{n_*}^\top)} \\
&= \frac{\mathbf{1}_{m n_*}^\top\,\mathrm{vec}(\mathbf{R}_s^{-1} \mathbf{Y}^* \mathbf{R}_x^{-1})}
        {\mathbf{1}_{m n_*}^\top\,\mathrm{vec}(\mathbf{R}_s^{-1} \mathbf{1}_m \mathbf{1}_{n_*}^\top \mathbf{R}_x^{-1})},
\end{split}
\end{equation}

\begin{equation}
\begin{split}
\hat{\sigma}^2
&= \frac{1}{mn_*}\,
   \bigl(\mathrm{vec}(\mathbf{Y}^*) - \hat{\mu}\,\mathbf{1}_{mn_*}\bigr)^\top
   (\mathbf{R}_x^{-1} \otimes \mathbf{R}_s^{-1})
   \bigl(\mathrm{vec}(\mathbf{Y}^*) - \hat{\mu}\,\mathbf{1}_{mn_*}\bigr) \\
&= \frac{1}{mn_*}\,
   \bigl(\mathrm{vec}(\mathbf{Y}^*) - \hat{\mu}\,\mathbf{1}_{mn_*}\bigr)^\top
   \mathrm{vec}\!\left\{
      \mathbf{R}_s^{-1}\bigl(\mathbf{Y}^* - \hat{\mu}\,\mathbf{1}_m \mathbf{1}_{n_*}^\top\bigr)\mathbf{R}_x^{-1}
   \right\}.
\end{split}
\end{equation}

Let $\boldsymbol{\theta}=(\theta_{x,1:p},\,\theta_{s,1:q})^\top\in\mathbb{R}^{\,p+q}$ denote the vector of kernel hyperparameters (temporal and spatial length-scales). They can be estimated by \citet{Rasmussen2006}: 
\begin{equation}
\begin{split}
\hat{\boldsymbol{\theta}}
&= \arg\min_{\boldsymbol{\theta}}
   \left\{ mn_* \log \hat{\sigma}^2 + \log|\mathbf{R}| \right\} \\
&= \arg\min_{\boldsymbol{\theta}}
   \left\{ mn_* \log \hat{\sigma}^2
      + n_* \log|\mathbf{R}_s|
      + m \log|\mathbf{R}_x| \right\}.
\end{split}
\end{equation}

\subsubsection{Prediction}
\vspace{-9 pt}

Let \( \mathbf{x}^\dagger \in \mathbb{R}^p \) and \( \mathbf{s}^\dagger \in \mathbb{R}^q \) be the test inputs. The kernel vectors are:
\begin{align}
\mathbf{r}_x(\mathbf{x}^\dagger) &= \left\{ \exp\left( -\sum_{l=1}^p \frac{|x^*_{lu} - x^\dagger_l|}{\theta_{x,l}} \right) \right\}_{u=1}^{n_*}, \\
\mathbf{r}_s(\mathbf{s}^\dagger) 
&= \left\{ 
\left[ 1 + \sum_{g=1}^q 
\left( \frac{s_{ag} - s^\dagger_g}{\theta_{s,g}} \right)^2 
\right]^{-1} 
\right\}_{a=1}^m.
\end{align}
Then, the ordinary kriging (OK) predictor becomes: 
\begin{equation}
\begin{aligned}
\hat{y}(\mathbf{s}^\dagger, \mathbf{x}^\dagger)
&= \hat{\mu} + \big( \mathbf{r}_x(\mathbf{x}^\dagger) \otimes \mathbf{r}_s(\mathbf{s}^\dagger) \big)^\top 
   \big( \mathbf{R}_x^{-1} \otimes \mathbf{R}_s^{-1} \big) 
   \big( \mathrm{vec}(\mathbf{Y}^*) - \hat{\mu}\, \mathbf{1}_{mn_*} \big) \\[6pt]
&= \hat{\mu} + \big( \mathbf{r}_x(\mathbf{x}^\dagger)^\top \otimes \mathbf{r}_s(\mathbf{s}^\dagger)^\top \big) 
   \, \mathrm{vec}\!\Big( \mathbf{R}_s^{-1} \big( \mathbf{Y}^* - \hat{\mu}\, \mathbf{1}_m \mathbf{1}_{n_*}^\top \big) \mathbf{R}_x^{-1} \Big) \\[6pt]
&= \hat{\mu} + \mathbf{r}_s(\mathbf{s}^\dagger)^\top \mathbf{R}_s^{-1} 
   \big( \mathbf{Y}^* - \hat{\mu}\, \mathbf{1}_m \mathbf{1}_{n_*}^\top \big) 
   \mathbf{R}_x^{-1} \mathbf{r}_x(\mathbf{x}^\dagger).
\end{aligned}
\end{equation}
%This corresponds to ordinary kriging (OK) under a constant mean and separable covariance model \citep{Olea2000}. 
However, OK suffers from the mean reversion issue in regions with sparse data, especially when modeling nonstationary functions such as power curves. %This phenomenon is illustrated in Figure~\ref{fig:lk_ok_comparison}, where the OK predictor (red line) tends to revert back to the global mean in less dense regions. 
One way to overcome this issue is to add a mean function that captures the nominal speed-power relationship shown in Figure \ref{fig:power_curves}. However, this relationship is nonlinear and thus cannot be handled using a universal kriging model \citep{Cressie1991}.  Moreover, the speed-power relationship can change with respect to the turbines and therefore, require additional modeling to capture the relationship between the parameters in the nonlinear model and the terrain characteristics.

% \begin{figure}[htbp]
% \centering
% \includegraphics[width=0.7\textwidth]{CKD-20250830-SpatioTemporalGP/lk_ok.pdf}
% \caption{Comparison of prediction methods for a representative turbine. 
% \textbf{Red:} Ordinary Kriging (OK) reverts toward the global mean in data-sparse regions (low/high wind speeds).}
% \textbf{Blue:} Limit Kriging (LK) maintains the full range of variation, including in regions with sparse training data. 
% \label{fig:lk_ok_comparison}
% \end{figure}
A much simpler approach to avoid mean reversion is to use the limit kriging (LK) proposed in \cite{Joseph2006}. The LK predictor is given by
\begin{equation}
\begin{aligned}
\hat{y}(\mathbf{s}^\dagger, \mathbf{x}^\dagger)
&= \frac{\big( \mathbf{r}_x(\mathbf{x}^\dagger) \otimes \mathbf{r}_s(\mathbf{s}^\dagger) \big)^\top 
   \big( \mathbf{R}_x^{-1} \otimes \mathbf{R}_s^{-1} \big) 
   \mathrm{vec}(\mathbf{Y}^*)}{\big( \mathbf{r}_x(\mathbf{x}^\dagger) \otimes \mathbf{r}_s(\mathbf{s}^\dagger) \big)^\top 
   \big(\mathbf{R}_x^{-1} \otimes \mathbf{R}_s^{-1}\big)  
  \mathbf{1}_{mn_*}}   \\[6pt]
&= \frac{\mathbf{r}_s(\mathbf{s}^\dagger)^\top \mathbf{R}_s^{-1} 
   \mathbf{Y}^*  \mathbf{R}_x^{-1} \mathbf{r}_x(\mathbf{x}^\dagger)}{\mathbf{r}_s(\mathbf{s}^\dagger)^\top \mathbf{R}_s^{-1} \mathbf{1}_{m}\mathbf{1}_{n_*}^\top
     \mathbf{R}_x^{-1} \mathbf{r}_x(\mathbf{x}^\dagger)}.
\end{aligned}
\end{equation}
%The LK predictor is shown as a blue dashed line in Figure~\ref{fig:lk_ok_comparison}, whose predictions remain robust even in regions with sparse data. Figure~\ref{fig:lk_ok_comparison} illustrates the case where wind speed is the only covariate.% 
%When the model is extended to multi-dimensional inputs and the spatio-temporal setting, the shortcomings of OK become even more pronounced. 
A numerical comparison between OK and LK will be provided in the next section.
%To extend limit kriging to the spatio-temporal setting, we define:
%\begin{equation}
%\mathbf{A} = \mathbf{R}_s^{-1} \mathbf{1}_m \mathbf{1}_{n_*}^\top %\mathbf{R}_x^{-1}, \quad
%\mathbf{B} = \mathbf{R}_s^{-1} \mathbf{Y}^* \mathbf{R}_x^{-1}.
%\end{equation}
%Then the predictive mean follows:
%\begin{equation}
%\hat{y}(\mathbf{s}^\dagger, \mathbf{x}^\dagger) = %\frac{\mathbf{r}_s(\mathbf{s}^\dagger)^\top \mathbf{B} \, %\mathbf{r}_x(\mathbf{x}^\dagger)}{\mathbf{r}_s(\mathbf{s}^\dagger)^\top \mathbf{A} \, {\mathbf{r}_x(\mathbf{x}^\dagger)}}.
%\end{equation}
To quantify predictive uncertainty at a test location \( (\mathbf{s}^\dagger, \mathbf{x}^\dagger) \), we use the Mean Squared Prediction Error (MSPE) formula \citep{Joseph2006}:
\begin{equation}
V(\mathbf{s}^\dagger, \mathbf{x}^\dagger)
= \operatorname{Var}\!\big(\hat{y}(\mathbf{s}^\dagger, \mathbf{x}^\dagger)\big)
= \hat{\sigma}^2 \left[ 1 - \alpha + \alpha \left( \frac{1 - \beta}{\beta} \right)^2 \right]
+ \eta_{t^\dagger}^{(i^\dagger)} .
\end{equation}
where
\begin{align}
\alpha &= \mathbf{r}_x(\mathbf{x}^\dagger)^\top \mathbf{R}_x^{-1} \mathbf{r}_x(\mathbf{x}^\dagger)\mathbf{r}_s(\mathbf{s}^\dagger)^\top \mathbf{R}_s^{-1} \mathbf{r}_s(\mathbf{s}^\dagger), \\
\beta      &= \mathbf{r}_s(\mathbf{s}^\dagger)^\top \mathbf{R}_s^{-1} \mathbf{1}_m \mathbf{1}_{n_*}^\top \mathbf{R}_x^{-1} \, \mathbf{r}_x(\mathbf{x}^\dagger).
\end{align}
The observation-noise variance, $\eta_{t^\dagger}^{(i^\dagger)}$, is obtained by Algorithm~\ref{alg:10nn_sigma2_notrans} corresponding to
the closest temporal and spatial locations to the test point:
\begin{equation}
t^\dagger
=
\argmin_{t=1,\ldots,n_*}
\sum_{l=1}^p
\left( x^*_{lt} - x^\dagger_l \right)^2,
\qquad
i^\dagger
=
\argmin_{i=1,\ldots,m}
\sum_{r=1}^q
\left( s^{(i)}_{r} - s^\dagger_{r} \right)^2 .
\end{equation}

One limitation of the proposed formulation is that, although it incorporates heteroskedastic observation noise $\eta_t^{(i)}$, it ignores the first-stage latent process uncertainty $\eta_{*,t}^{(i)}$ and therefore uses a noise-free covariance for the pseudo outputs. The main advantage of this simplified formulation is its conceptual simplicity and computational efficiency. The latent process variance $\eta_{*,t}^{(i)}$ can be incorporated into the covariance structure through some approximations. However, we did not find any improvement in the numerical results, and therefore, the details are presented in the Supplementary Material (S5).

\vspace{-9 pt}
\section{Numerical Experiments}
\label{sec:experiments}
\vspace{-9 pt}
The numerical experiments are designed to evaluate the performance of the proposed STGP model and demonstrate unique capabilities enabled by its terrain-incorporating structure.

\vspace{-9 pt}
\subsection{Experimental Setup}
\label{ssec:data_criteria}
\vspace{-9 pt}

We use 10-minute SCADA measurements from 66 turbines in an onshore wind farm over two calendar years (2017–2018). Each record includes timestamped values for wind speed, wind direction, ambient temperature, turbulence intensity, standard deviation of wind direction, and active power output. \citet{Prakash2024}, who performed a comprehensive sensitivity analysis on this dataset, found that using only wind speed and temperature as the temporal inputs produces the best-performing temporal-only power curve model for this data.  We tested the data with our model set up and found that their finding is still true.  We also use wind speed and temperature as temporal inputs and exclude other environmental covariates such as turbulence intensity and wind direction. We want to stress that which temporal covariates to include may vary depending on the specific wind farm; previous studies (for example, \cite{Lee2015a}) show that wind direction and humidity could also be significant. In addition to these temporal measurements, each turbine is associated with three static terrain descriptors: terrain slope, RIX, and ridge height. These terrain features capture spatial heterogeneity relevant to wind flow and turbine performance. 

Four turbines (IDs 47, 51, 53, and 61) were known to suffer from sensor drift in 2018. While their 2018 records are excluded from the evaluation, their 2017 data are retained for both training and testing.

Model performance is assessed using two standard metrics: Root Mean Squared Error (RMSE), which quantifies the accuracy of point predictions, and Negative Log Predictive Density (NLPD), which evaluates the calibration and sharpness of the full predictive distribution. Lower values for both metrics indicate superior performance. Let the test set be 
\(\bigl\{(\mathbf{x}^{\dagger}_i,\mathbf{s}^{\dagger}_i),\,y^{\dagger}_i\bigr\}_{i=1}^{n_{\text{test}}}\),
where \(\mathbf{x}^{\dagger}_i\) represents temporal covariates (wind speed and temperature), \(\mathbf{s}^{\dagger}_i\) denotes terrain features, and \(y^{\dagger}_i\) is the observed power output. Let \(\hat{y}(\mathbf{x}^{\dagger}_i,\mathbf{s}^{\dagger}_i)\) and \(V(\mathbf{x}^{\dagger}_i,\mathbf{s}^{\dagger}_i)\) denote the predictive mean and variance. The two metrics are defined as follows:

\begin{equation}
\label{eq:rmse_dagger}
\mathrm{RMSE} \;=\;
\sqrt{\frac{1}{n_{\text{test}}}\,
       \sum_{i=1}^{n_{\text{test}}}
       \bigl\{y^{\dagger}_i-\hat{y}(\mathbf{x}^{\dagger}_i,\mathbf{s}^{\dagger}_i)\bigr\}^{2}},
\end{equation}

\begin{equation}
\label{eq:nlpd_dagger}
\mathrm{NLPD} \;=\;
\frac{1}{2\,n_{\text{test}}}\,
\sum_{i=1}^{n_{\text{test}}}
\left[
\frac{\bigl\{y^{\dagger}_i-\hat{y}(\mathbf{x}^{\dagger}_i,\mathbf{s}^{\dagger}_i)\bigr\}^{2}}
     {V(\mathbf{x}^{\dagger}_i,\mathbf{s}^{\dagger}_i)}
\;+\;
\log\!\bigl(2\pi\,V(\mathbf{x}^{\dagger}_i,\mathbf{s}^{\dagger}_i)\bigr)
\right].
\end{equation}

\subsection{Experiment 1: Leave-one-turbine-out Model Generalization Test}
\label{ssec:exp1}
\vspace{-9 pt}

The first set of experiments focuses on evaluating the model's predictive accuracy. This experiment evaluates each model’s ability to generalize spatially to an unseen turbine using a leave-one-turbine-out (LOTO) strategy. In each fold, one turbine is held out entirely for testing, while the remaining 65 turbines are used for training.

\begin{figure}[htbp]
\centering
\begin{minipage}{0.48\textwidth}
  \centering
  \includegraphics[width=\textwidth]{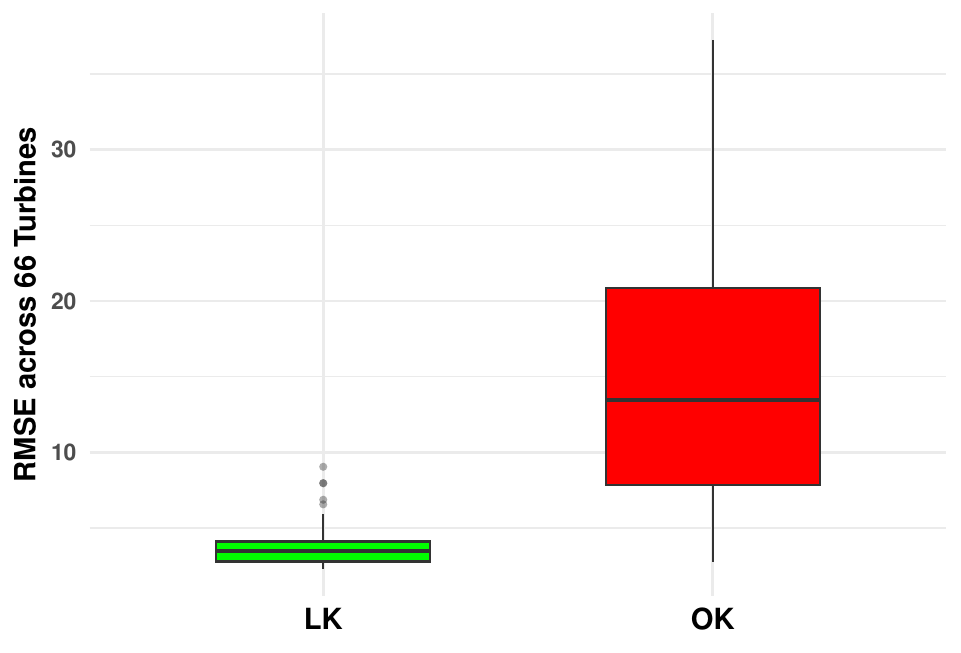}
\end{minipage}\hfill
\begin{minipage}{0.48\textwidth}
  \centering
  \includegraphics[width=\textwidth]{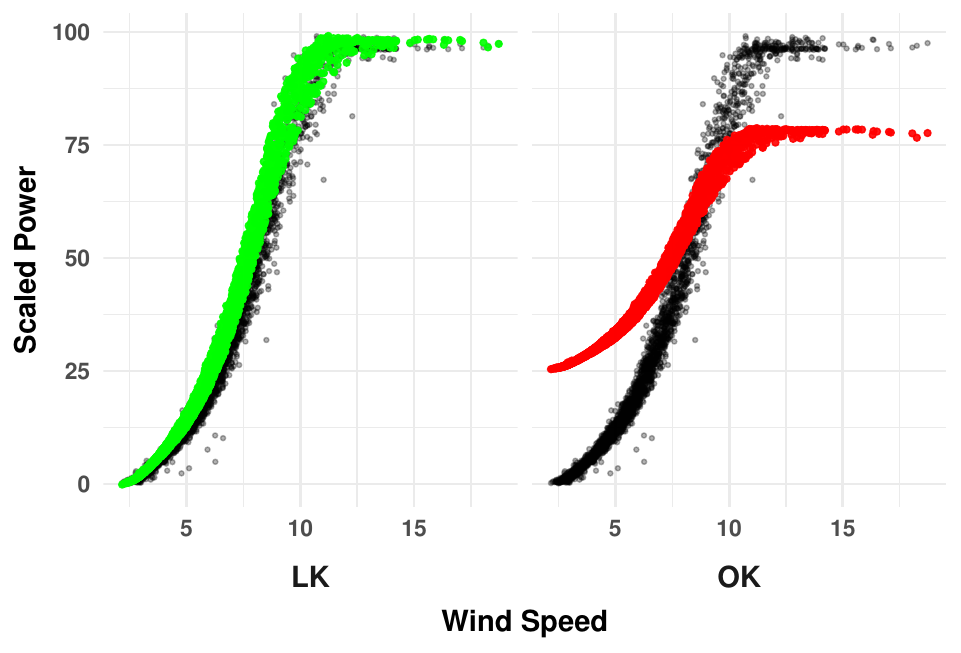}
\end{minipage}
\caption{Comparison of LK and OK. 
\textbf{Left:} RMSE comparison for 66 turbines in the LOTO experiment. 
  OK yields higher errors for most turbines. 
  LK achieves superior accuracy with lower variability.
\textbf{Right:} Power predictions for Turbine~\#60. 
The OK prediction (red) is pulled towards the global mean, whereas LK (green) provides a much better fit.}
\label{fig:lk_ok_combined}
\end{figure}

Figure~\ref{fig:lk_ok_combined} compares LK and OK under the LOTO setting with both training and testing on 2017 data. Both the methods employ separable covariance functions with the same kernels and parameters. While OK performs reasonably well for a few turbines, in most turbines, its RMSE is substantially larger than that of LK. An example is shown on the right panel, which clearly shows the mean reversion issue of OK. In the remainder of this paper, we focus exclusively on the LK-based STGP.

We evaluated the proposed STGP model against baseline methods under two training–testing setups designed to assess both approximation and forecasting capabilities: (i) training and testing on data from the same year (2017), and (ii) training on 2017 data and predicting on 2018 data. The second setup is particularly relevant for real-world applications such as site planning, where only historical turbine data are available at deployment time, and it provides a direct test of the model’s ability to generalize temporally from retrospective to forward-looking scenarios.

The IEC binning method, widely used in the wind industry, serves as a nonparametric baseline using only wind speed to define the power curve.  To compare with general machine learning method, we choose the top three methods identified in a recent wind turbine power curve challenge~\citep{Barber2024}: a multi-layer neural network(NN)-based, a GP-based, and a XGBoost-based power curve. The GP-based power curve reported in \citet{Barber2024} uses the tempGP~\citep{Prakash2023}, but tempGP runs too slow to handle the data size in this study. We here substitute tempGP in the GP-based power curve with twinGP for computational feasibility.  Following a reviewer’s recommendation, we also benchmarked our method against the PyTorch implementation of a Bayesian neural network (Bayesian NN) provided on the GitHub page of \citep{lee2022}.

Bayesian NN, BHM \citep{Prakash2024}, and our proposed method, STGP, incorporate both dynamic (temporal) and static (terrain) features. The other three machine learning models (multi-layer NN, XGBoost, twinGP) can also use both spatial and temporal inputs. In doing so, the static terrain features and dynamic environmental variables are concatenated into a single input vector, namely that static terrain variables are simply repeated for all observations for a given turbine and used as regular predictors together with the dynamic variables. Our investigation shows that compared to the same models that use dynamic variables alone, including terrain features in these three machine learning models consistently leads to a deterioration in predictive performance. For this reason, we decide to present the results from the respective model using dynamic variables alone in the main paper but include the aforementioned full comparison in the Supplementary Material (S1).

Table \ref{tab:exp1_2017} shows performance on 2017 test data, and Table \ref{tab:exp1_2018} on 2018. NLPD is not available for multi-layer NN, XGBoost, and IEC binning as the outcomes of these models do not include predictive uncertainty. Runtime includes both model estimation and prediction. The pseudo-inputs and pseudo-outputs are each computed once across all turbines; for fairness, the total cost of these shared preprocessing steps is divided by 66 and added to the per-fold training and prediction time of the spatio-temporal GP model.

\begin{table}[h!]
\centering\small
\caption{Experiment 1 (2017 test): RMSE, RMSE increase, NLPD, and runtime averaged over 66 turbines. Runtime includes both model estimation and prediction.}
\label{tab:exp1_2017}
\begin{tabular}{lcccc}
\toprule
\textbf{Method} & avg. RMSE & \%RMSE increase over best & avg. NLPD & avg. runtime (sec) \\
\midrule
STGP (ours)   & \textbf{3.73} & — & \textbf{2.74} & 197.92 \\
XGBoost        & 3.91 & 4.8\% & — & 3.43 \\
Multi-layer NN & 3.93 & 5.4\% & — &  219.21 \\
TwinGP         & 3.98 & 6.7\% & 3.20 & 129.97 \\
BHM        & 4.03 & 8.0\% & 2.91 & 17,808.88 \\
Bayesian NN   & 4.44 & 19.0\% & 2.92 & 161.85\\
IEC Binning        & 4.57 & 22.5\% & — & 1.79 \\
\bottomrule
\end{tabular}
\end{table}

\begin{table}[h!]
\centering\small
\caption{Experiment 1 (2018 test): RMSE, RMSE increase, NLPD, and runtime averaged over 62 turbines. Runtime includes both model estimation and prediction.}
\label{tab:exp1_2018}
\begin{tabular}{lcccc}
\toprule
\textbf{Method} & avg. RMSE & \% RMSE increase over best & avg. NLPD & avg. runtime (sec) \\
\midrule
STGP (ours)   & \textbf{3.97} & — & \textbf{2.79} & 197.89 \\
Multi-layer NN & 4.13 & 4.0\% & — & 218.98 \\
XGBoost        & 4.14 & 4.3\% & — & 3.68 \\
BHM        & 4.15 & 4.5\% & 2.95 & 17,808.75 \\
TwinGP         & 4.44 & 11.8\% & 3.37 & 121.47 \\
IEC Binning        & 4.62 & 16.4\% & — & 1.79 \\
Bayesian NN   & 4.69 & 18.1\% & 2.96 & 159.89\\
\bottomrule
\end{tabular}
\end{table}

STGP achieves the lowest RMSE across all methods for both years and the lowest NLPD among those whose NLPDs are available. STGP runs under 4 minutes and remains computationally practical for offline deployment. In terms of RMSE, XGboost and multi-layer NN are in the second best group. BHM's performance varies in the two experiments but it is the slowest. In our implementation, BHM attains an RMSE of 4.03 in the same-year testing, which is similar to but slightly better than the results reported by \citet{Prakash2024} themselves. Since TwinGP is used as a model element in STGP, let us take a closer look at TwinGP.  In terms of RMSE, STGP is 7–12\% better than TwinGP in the two evaluations. In terms of NLPD, STGP has an obvious advantage over TwinGP. Overall, the results confirm that explicitly incorporating terrain features enhances the model’s ability to generalize to unseen turbines. 

\vspace{-9 pt}
\subsection{Experiment 2: Quantifying Environmental and Terrain Importance}
\label{sec:sensitivity}
\vspace{-9 pt}

In the second experiment, we use feature attribution using Accumulated Local Effect (ALE) plots \citep{Apley2016}, which quantify the influence of spatial and temporal inputs on predicted power. This shows which covariates drive power variation, for interpretability. 

ALE plots measure the average change in a model’s prediction as a feature varies locally, offering more robust insights than partial dependence by avoiding extrapolation in sparse regions. In our setting, the STGP serves as black-box model. Because the ALE package does not handle inputs of different sizes, we computed plots by fixing either spatial or temporal inputs. Temporal features (wind speed, temperature) are evaluated at fixed spatial locations (turbines), and spatial features (slope, RIX, ridge height) at fixed temporal indices. 

\begin{figure}[htbp]
\vspace{10pt}
\centering
% --- First Row: Temporal Variables ---
\begin{minipage}{0.5\textwidth}
\centering
\includegraphics[width=\linewidth]{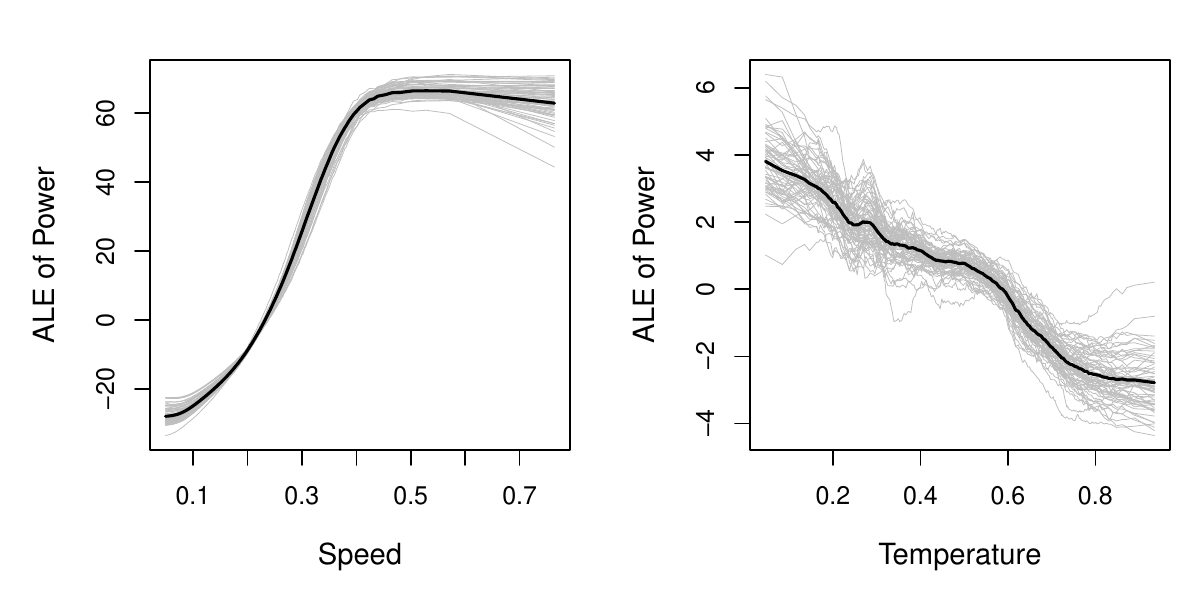}\\
%\scriptsize ALE: Wind Speed and Temperature
\end{minipage}

\vspace{.5em}

% --- Second Row: Spatial Variables ---
\begin{minipage}{0.66\textwidth}
\centering
\includegraphics[width=\linewidth]{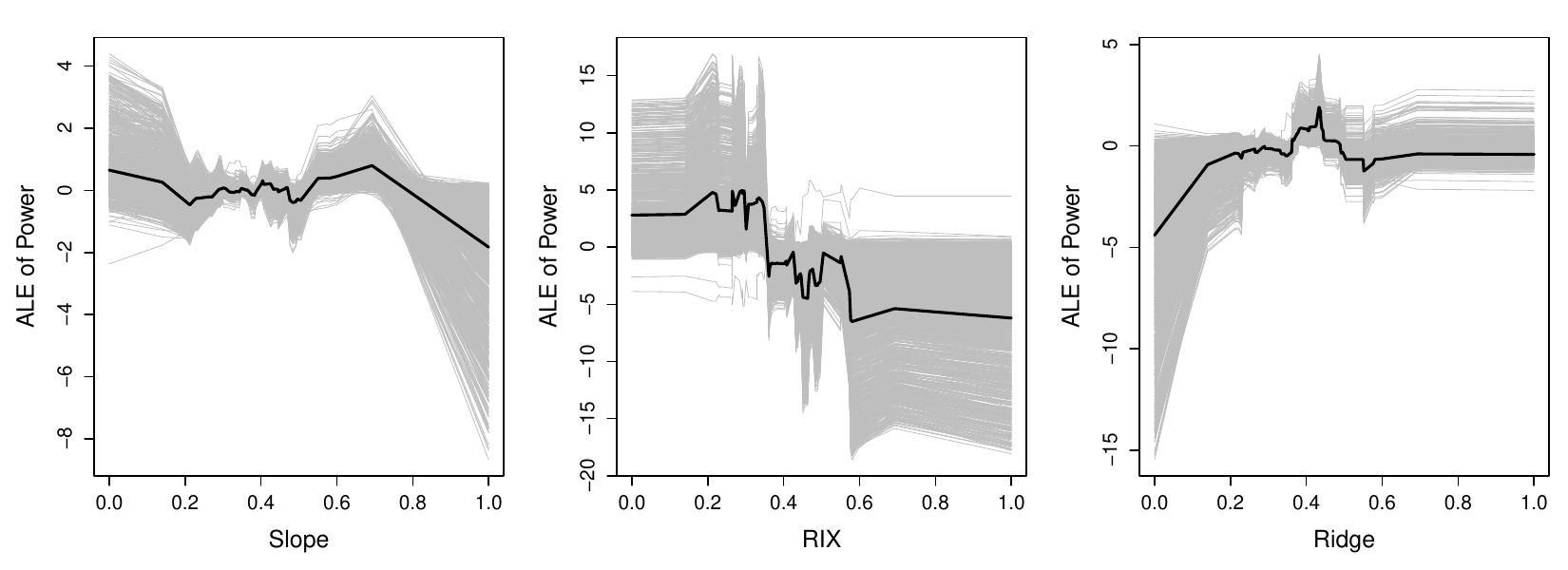}\\
%\scriptsize ALE: Slope, RIX, and Ridge
\end{minipage}
\caption{ALE plots for temporal and spatial features. Gray lines represent conditional ALEs computed by fixing spatial inputs (for temporal features) or temporal inputs (for spatial features). The black line shows the average ALE across all conditions.}

\label{fig:ale_plots}
\end{figure}

Each gray curve in Figure~\ref{fig:ale_plots} corresponds to an ALE under a fixed input, while the black curve represents their mean, highlighting the typical effect of each feature.

The ALE curves highlight the relative importance of the predictors. Wind Speed shows a steep power increase as values rise to 0.5 and then saturates, mimicking physical turbine behavior. The effect range spans nearly 94.3 units (range of the black line), confirming wind speed as the dominant predictor. Temperature has a monotonic effect, with predicted power declining by 6.6 units, consistent with reduced air density at higher temperatures. Slope contributes the least (~2.6 units), with a weakly negative trend and local fluctuations suggesting modest sensitivity to terrain inclination. RIX yields the most pronounced spatial effect, with a nearly 11.5 unit drop as ruggedness increases, indicating that rougher terrain disrupts airflow and diminishes performance. Finally, Ridge Height has a smaller effect, with a 6.3 unit increase for smaller ridge values that stabilizes at larger values.

For spatial features, higher ridge values and small RIX levels are associated with increased power generation, while slope peaks at moderate to high values. Results confirm that terrain features, especially RIX, significantly influence power and underscore the importance of incorporating terrain variables in turbine performance modeling. We believe this is one of the first times these terrain effects are quantified in actual turbine operations.

\vspace{-9 pt}
\subsection{Experiment 3: Extrapolation to Distant Turbines}
\label{sec:exp4}
\vspace{-9 pt}

The third experiment evaluates the model's ability to extrapolate across geographic domains. In some wind energy applications, the objective is to estimate the potential power generation of turbines in a distant wind farm using a model trained on data from another farm. To simulate this setting, we divide the turbines into two groups, where the smaller group is geographically distant from the larger one. The model is trained on the larger group and evaluated on the smaller group. We then compare the accuracy of terrain-aware models such as Bayesian NN, BHM, and the proposed STGP with terrain-agnostic alternatives.

\begin{figure}[ht]
\centering
\includegraphics[width=0.6\textwidth]{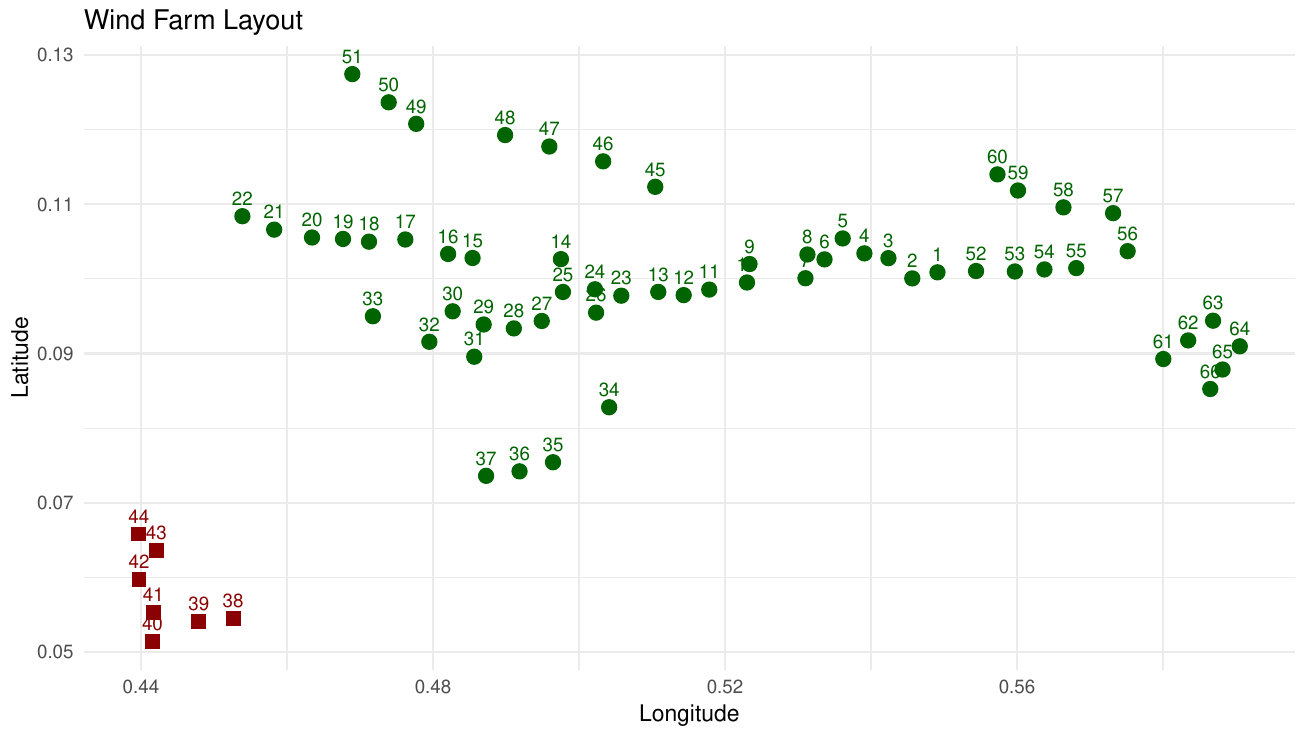}
\caption{Spatial distribution of turbines across the farm. Distant turbines (squares) are treated as potential locations for future turbines and are used for testing, while the remaining turbines represent the existing farm with historical data used for model training.}
\label{fig:map}
\end{figure}

This application is relevant for planning new wind farms, where temporal and spatial measurements can be collected at several candidate locations. Predictions from a model trained on data from an existing wind farm can then help identify locations with higher expected power generation. Figure~\ref{fig:map} shows the layout of the two groups, where turbines 38 to 44 (squares) form the smaller group with 7 turbines used for testing, while the remaining turbines (circles) are used for training.

Based on the suggestion of one of the reviewers, we investigated an additional 
alternative to the proposed temporal alignment strategy by discretizing the continuous 
inputs $x_t$ through binning wind speed and temperature values at a fixed resolution, 
thereby reducing the effective sample size to a computationally tractable level for 
standard GP fitting. Rounding these covariates to the nearest integer ($\pm 0.5$ bin 
width) collapsed the roughly 2.4 million training observations to 41,569 unique locations; however, this remains computationally prohibitive for full GP inference. Progressively 
increasing the bin width to $\pm 1.5$ further reduced the dataset to 5,884 
representative samples, a size tractable to fitting via the \texttt{hetGP} 
package which can efficiently handle replicated observations \citep{Binois2021}.

\begin{table}[h!]
\centering\small
\caption{Experiment 3: RMSE for extrapolation to geographically distant turbines}
\label{tab:exp3}
\begin{tabular}{lcccc}
\toprule
\multirow{2}{*}{Method} & \multicolumn{2}{c}{Features: $x$} & \multicolumn{2}{c}{Features: $x+s$} \\
\cmidrule(lr){2-3} \cmidrule(lr){4-5}
 & RMSE & \% Increase & RMSE & \% Increase \\
\midrule
STGP (ours)    & --   & --      & \textbf{3.40} & --      \\
XGBoost        & 3.63 & 6.8\%   & 4.01          & 17.9\%  \\
Multi-layer NN & 3.62 & 6.5\%   & 4.29          & 26.2\%  \\
BHM            & --   & --      & 4.62          & 35.9\%  \\
Bayesian NN    & 3.64 & 7.1\%   & 4.81          & 41.5\%  \\
TwinGP         & 3.74 & 10.0\%  & 4.74          & 39.4\%  \\
IEC Binning    & 4.28 & 25.9\%  & --            & --      \\
Binning-hetGP  & 4.49 & 32.1\%  & 4.62          & 35.9\%  \\
\bottomrule
\end{tabular}
\end{table}

Table~\ref{tab:exp3} shows that STGP outperforms all baselines in the extrapolation setting. By leveraging learned relationships between terrain and wind behavior, STGP predicts effectively at unseen, geographically distant locations. For most methods, if applicable, we implement two versions: one using only $x$ as features, and one using both $x$ and $s$, where the terrain vector is replicated across all data points from the same turbine. In almost all cases, the $x+s$ version shows a larger loss in accuracy, suggesting that naively appending terrain features is insufficient. Bayesian NN suffers the most from this simple addition, while the multi-layer NN without terrain, despite underperforming STGP, maintains a relatively smaller accuracy loss compared to the other baselines. The BHM model relies on a parametric link between terrain and the response, which proves limiting in this setting — whereas BHM was only 5--8\% worse than STGP in Experiment 1 (Section~\ref{ssec:exp1}), the gap widens to 35.9\% here. Although Bayesian NN also incorporates terrain information, the results suggest that terrain variables require more specialized modeling than simply including them alongside temporal features. The Binning-hetGP models produced relatively large predictive errors, suggesting that naive discretization strategies for resolving temporal misalignment fall substantially short of the proposed STGP.

This experiment demonstrates the importance of incorporating non-parametric domain-specific terrain knowledge for reliable extrapolation. It highlights STGP’s unique capacity to support informed wind farm planning in geographically distant locations.

\vspace{-9 pt}
\subsection{Experiment 4: Evaluating Turbine Underperformance with Terrain-Aware Modeling}
\label{sec:underperformance}
\vspace{-9 pt}

The final set of experiments examines the model’s utility for turbine-level diagnostics, a natural byproduct of its terrain-aware formulation. We test whether the model can identify underperforming turbines by separating performance loss due to terrain from that due to turbine inefficiency. Compared with a terrain-agnostic alternative, the terrain-aware model reduces false alarms and improves maintenance decision-making.

Our aim is not to provide a definitive measure of efficiency but to show how the STGP can decompose observed deficits into terrain-driven and turbine-specific components, yielding more interpretable diagnostics. A key strength is its ability to separate poor performance from environmental disadvantages. To illustrate, we evaluate all turbines under identical environmental inputs \( \mathbf{X}^* \), so variation arises only from turbine-specific effects, terrain, or noise. The GP model estimates and removes the terrain contribution.

We begin with Turbine 54 (T54), known for low output. Let \( P_{\text{T54}} \) be its observed power under \( \mathbf{X}^* \) and \( \bar{P}_{\text{others}} \) the peer average. A naive gap is 
\(
\text{NaiveUnderperf} = \bar{P}_{\text{others}} - P_{\text{T54}},
\) 
which ignores terrain. Using the GP trained without T54, we predict \( \hat{P}_{\text{T54}}^{\mathbf{s}_{54}} \) (T54’s terrain) and \( \hat{P}_{\text{T54}}^{\bar{\mathbf{s}}} \) (average peer terrain). The terrain effect is 
\(
\text{TerrainEffect} = \hat{P}_{\text{T54}}^{\bar{\mathbf{s}}} - \hat{P}_{\text{T54}}^{\mathbf{s}_{54}},
\) 
and the adjusted deficit is 
\(
\text{AdjustedUnderperf} = (\bar{P}_{\text{others}} - P_{\text{T54}}) - \text{TerrainEffect}.
\)
For T54, the normalized gaps are 
\(
\text{NaiveUnderperf}/\bar{P}_{\text{others}} \approx 18.9\% 
\) and 
\(
\text{AdjustedUnderperf}/\bar{P}_{\text{others}} \approx 13.2\%,
\) 
showing that 5.6\% of the shortfall stems from terrain while 13.2\% reflects inefficiency.

\begin{figure}[ht]
\centering
\includegraphics[width=0.6\textwidth]{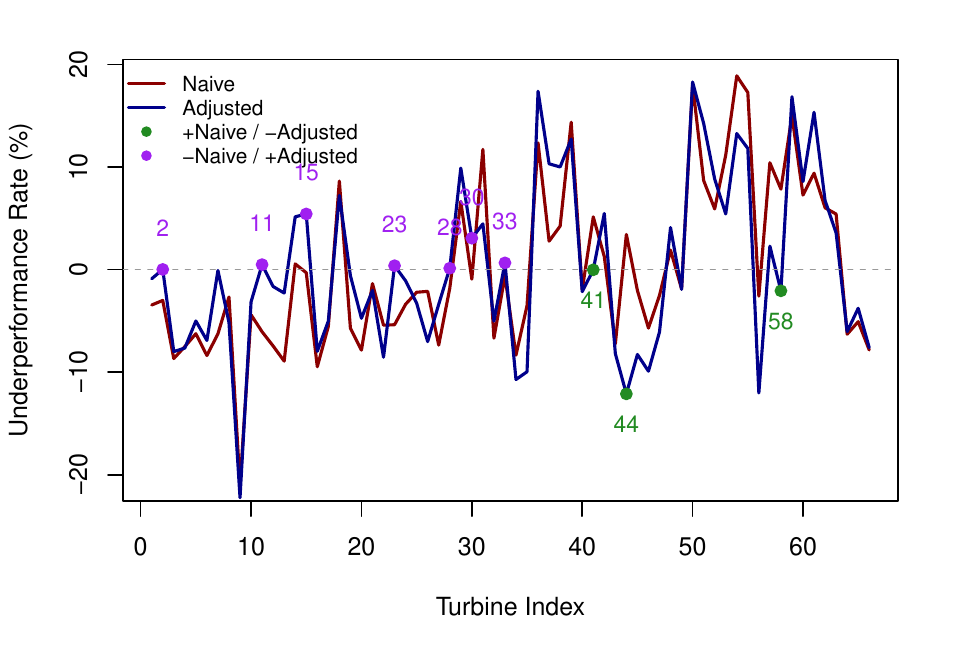}
\caption{Naive (red) and adjusted (blue) underperformance rate (\%) for each turbine. Turbines with sign reversal between the two metrics are labeled.}
\label{fig:underperf_comparison}
\end{figure}

We extend this analysis to all turbines. Figure~\ref{fig:underperf_comparison} compares naive (red) and adjusted (blue) rates. Turbines with both positive are genuinely underperforming; both negative indicates overperformance; positive naive but negative adjusted implies terrain-driven deficit; and negative naive but positive adjusted means masked underperformance.

Examples include turbines 41, 44, and 58 (terrain-driven deficits) and 2, 11, 15, 23, 28, 30, and 33 (masked underperformance). For some, such as 44 and 58, the terrain effect critically alters interpretation. This experiment demonstrates how terrain-aware modeling provides a transparent way to decompose performance, avoiding false alerts and uncovering inefficiencies that peer comparisons alone could miss.

%\addtolength{\textheight}{-.2in}%

\vspace{-9 pt}
\section{Conclusion}
\label{sec:conclusion}
\vspace{-9 pt}

This work introduces a scalable spatio-temporal Gaussian process framework for modeling wind turbine power curves, integrating dynamic environmental variables with static terrain features. The proposed method reduces the large unstructured temporal data into small data on a shared temporal grid which enables efficient prediction and uncertainty quantification through the use of Kronecker products. We evaluated the proposed model using leave-one-turbine-out cross-validation on a real-world dataset comprising 66 turbines over two years (2017–2018). Across both same-year and out-of-year forecasting scenarios, our model outperformed benchmark methods like TwinGP, multi-layer NN, Bayesian NN, XGBoost, BHM, and IEC binning in terms of root mean squared error.

Accumulated Local Effects analysis further highlighted the dominant role of wind speed and quantified the contributions of terrain characteristics, with ruggedness identified as the most influential static variable. Under geographical extrapolation settings, our method showed more robust performance compared to alternatives, validating its generalizability for wind farm planning tasks.
Beyond predictive performance, the model enables insightful diagnostics. We demonstrated its utility in turbine performance benchmarking and showed how it distinguishes true underperformance from terrain-induced variation. 

In summary, our approach provides a unified, interpretable, and accurate modeling framework for wind power prediction and turbine evaluation. Future work may focus on developing more robust designs for performance quantification, applying the method to anomaly detection in operational data, and improving computational efficiency, particularly enabling adaptive selection of additional temporal support points after initial reduction.

\vspace{-9 pt}
\section*{Supplementary Material}
\spacingset{1.8}
\vspace{-9 pt}

\noindent \textbf{Supplementary Material:}\label{Appendices} A separate PDF includes S1 (Treating Terrain Features Like Extra Covariates), S2 (top-down terrain illustration), S3 (support points), S4 (testing different covariance functions), and S5 (STGP Including Process Noise of Pseudo-Outputs).

\noindent \textbf{Data Availability Statement:}\label{data-availability-statement}
\label{Datasets}
The dataset and computer code to reproduce key results are publicly available at the \href{https://github.com/anonymous-stgp-terrain/STGP-Terrain}{GitHub page}.

\vspace{-9 pt}
\section*{Disclosure statement}\label{disclosure-statement}
\vspace{-9 pt}

The authors report there are no competing interests to declare.

\vspace{-9 pt}
\section*{Funding}\label{funding}
\vspace{-9 pt}

		\if0\blind
{

YD’s and AC’s research was sponsored by the Ocean Energy Safety Institute Consortium (OESIC) through a grant from the U.S.\ Department of the Interior, Bureau of Safety and Environmental Enforcement (BSEE), and the U.S.\ Department of Energy (DOE) and was accomplished under Agreement Number E21AC00000. The views and conclusions contained in this document are those of the authors and should not be interpreted as representing the opinions or policies of the U.S.\ Government. Mention of trade names or commercial products does not constitute their endorsement by the U.S.\ Government. 
}\fi
		\if1\blind
{
Funding acknowledge is purposely removed for double-blind review.
}\fi

\vspace{6 pt}

\spacingset{1}
\bibliographystyle{apalike}

\bibliography{bibliography}

\end{document}